# A comparative investigation of thickness measurements of ultra-thin water films by scanning probe techniques

A. Opitz [a]

Institut für Physik, Universität Augsburg, 86135 Augsburg, Germany

Institut für Physik and Institut für Mikro- und Nanotechnologien, Technische Universität Ilmenau, PF 100565, 98694 Ilmenau, Germany

M. Scherge [b]

IAVF Antriebstechnik AG, Im Schlehert 32, 76187 Karlsruhe, Germany

S.I.-U Ahmed, and J.A. Schaefer

Institut für Physik and Institut für Mikro- und Nanotechnologien, Technische Universität Ilmenau, PF 100565, 98694 Ilmenau, Germany

---

[a] Email address: andreas.opitz@physik.uni-augsburg.de
[b] Present Address: Fraunhofer Institut für Werkstoffmechanik, Wöhlerstr. 11, 79108 Freiburg, Germany

ABSTRACT

The reliable operation of micro and nanomechanical devices necessitates a precise knowledge of the water film thickness present on the surfaces of these devices with accuracy in the nm range. In this work, the thickness of an ultra-thin water film was measured by distance tunnelling spectroscopy and distance dynamic force spectroscopy during desorption in an ultra-high vacuum system, from about 2.5 nm up to complete desorption at $10^{-8}$ mbar. The tunnelling current and the amplitude of vibration and the normal force were detected as a function of the probe-sample distance. In these experiments, a direct comparison of both methods was possible. It was determined that dynamic force spectroscopy provides the most accurate values. The previously reported tunnelling spectroscopy, which requires the application of significantly high voltages generally leads to values that are 25 times higher than values determined by dynamic force spectroscopy.

KEYWORDS

Scanning tunnelling microscopy (STM), scanning force microscopy (SFM), distance tunnelling spectroscopy, distance dynamic force spectroscopy, water film thickness, desorption

I. INTRODUCTION

An ultra-thin water layer whose thickness depends on the exact environmental parameters covers most surfaces exposed to ambient environments. Numerous reports have shown that this ultra-thin water film has considerable influence on the operation and reliability of microelectromechanical systems (MEMS), especially MEMS consisting of moving parts in contact due to capillary action [1-4]. This capillarity causes parts to adhere to one another and leads to operational instabilities and catastrophic failure of the device. The same mechanism causes failure sometimes even during device fabrication. Given the current popularity of silicon, with a native oxide layer in air that is hydrophilic, a precise knowledge about the thickness of the ultra-thin water film under various environmental conditions is necessary. Such knowledge will help in the design and fabrication of micro systems with better performance characteristics. It is easy to deduce that this information will become even more critical for future nanoelectromechanical devices (NEMS).

Besides moving MEMS and the upcoming NEMS, developments in nano-scale biological imaging also demand nm accuracy. Here the water film plays a significant role in the correct imaging of biological objects such as membranes, bacteria and viruses. Recently insulating membranes [5] and viruses [6] were mapped with a thin water film.

Previous measurements of water film thickness were investigated by scanning probe microscopy (SPM). During imaging the surface the water droplets were visible with dynamic scanning force microscopy [7], polarization force microscopy [8] and scanning tunnelling microscopy [9]. In contact mode scanning force microscope, a water droplet is not measurable because there is no contrast [7]. Also, the tip penetrates the water film. The water droplet height was measured by dynamic scanning force microscopy [7] on gold, mica and graphite.

This study reported that water droplets on graphite have a height of 2 nm at 35% RH and a height of 5 nm height at 95% RH. Water islands were measured on gold to have a molecular height of 0.2 nm, while 10 nm high water droplets was mapped with scanning tunnelling microscopy [9]. These methods are only capable of measuring water islands rather than completely covered surfaces because they utilize the edges of water droplets for obtaining a height contrast. Completely covered surfaces can be measured with distance spectroscopy methods. A detailed study using distance tunnelling spectroscopy reported a water film thickness on gold of more than 10 nm for a relative humidity above 35% RH and a thickness on titanium that was more than 50 nm for a relative humidity above 20% RH [9]. Compared to microscopic studies on water islands, such values are relatively high.

This paper investigates the measurement of the thicknesses of ultra-thin water films using two kinds of distance spectroscopy: First, distance tunnelling spectroscopy (DTS) reported previously [9] is used and secondly, another method - distance dynamic force spectroscopy (DDFS), is introduced. Both methods are suited for applications involving completely covered water surfaces. Samples fabricated from silicon wafers were used because this is an often-used material in microsystems. Native oxide covered silicon and hydrogen terminated silicon were measured to compare hydrophilic and hydrophobic surfaces.

## II. EXPERIMENTAL SETUP

Ultra-thin water films with varying thicknesses were generated by pumping a UHV chamber containing the measurement system from air pressure down to $10^{-8}$ mbar with a combination of turbo molecular, ion and titanium pumps [10,11]. As a consequence of the pumping process the water vapour pressure in the UHV chamber continually decreases, which results in the desorption of the water film from an initial thickness value down to almost zero.

Hydrophilic and hydrophobic silicon samples were used for examination. Native oxide-covered silicon (100-surface, 1-10 $\Omega$cm, B-doped) was used as the hydrophilic sample [12]. For this sample the native oxide thickness is about 1.3 nm. All samples were first cleaned using ultrasonic assistance for 5 min in isopropanol and methanol. A hydrophobic surface was created by etching the silicon samples for 1 min in concentrated HF (40% v/v). This results in complete removal of the oxide layer and a hydrogen terminated silicon surface, which is hydrophobic [12]. Afterwards all samples were rinsed using bi-distilled water before insertion into the UHV system.

For water film thickness measurements with DTS, a tungsten tip was used as the probe. This tip was prepared by etching a tungsten wire in NaOH to obtain a sharp tip. It is clear that an oxide is present at the tip immediately after the etching process. Silicon cantilevers were used in thickness measurements with DDFS. The silicon cantilevers (with resonance frequencies of about 300 kHz) were covered also by a native oxide. Both probes used are hydrophilic due to the oxide layer present on them.

III. DISTANCE TUNNELLING SPECTROSCOPY

DTS is based on current-distance measurement performed using the scanning tunnelling microscope [9]. A typical curve is shown in Fig. 1. The current, *I,* is measured during approach of the tungsten tip towards the sample as a function of the distance between the tip and sample given by the piezo value, *z,* which is converted to a length value from a voltage value using standard calibration samples. Far away from the surface (range (a) in Fig. 1) the measured current is less then 10 pA.

This is defined as the zero noise of the current measurement. Upon further approach, at a certain value, z1, the current suddenly increases and maintains a stable value of 100 pA (range (b)). With further decrease of the tip-sample distance, the current increases linearly (range (c)) up to the highest measurable current (here 50 nA), at point $z_2 = 0$, at which point the tip ultimately contacts the sample. In the measurements shown here the water film thickness is defined by

$$d_{H2O,DTS} = \frac{z_1}{2} \qquad (1)$$

The factor of half takes into consideration the water film that is present on the sample and tip and assumes that the same amount of water is present on both surfaces. The voltage used for these measurements was 10 V, which is relatively high. This high voltage was necessary to initiate current flow through the insulating silicon oxide layer.

On hydrophobic silicon the same voltage of 10 V was used for comparative purposes. In these measurements $z_2$ is not the real contact point between tip and sample [9]. The tunnelling resistance at this point is the same as in the case of $SiO_2$, which is about 200 MΩ. This is limited by the current measurement of the setup. Upon further approach the current remains constant while the contact resistance decreases.

IV. DISTANCE DYNAMIC FORCE SPECTROSCOPY (DDFS)

Distance dynamic force spectroscopy (DDFS) is based on the measurement of amplitude, *a*, and normal force, $F_N$, of a cantilever vibrating perpendicular to the surface as a function of the tip-sample distance, $z$, during approach of vibrating tip towards the sample. A typical curve is shown in Fig. 2. Far from the surface, the cantilever vibrates with the fixed amplitude $a_0$ (range (a)). As the tip approaches the sample, the cantilever starts to interact with the water/sample system (point $z_1$). The amplitude begins to decrease. At point $z_2$ the jump into contact occurs. Upon further piezo movement, bending of cantilever in contact with the sample is seen (range (c) and (d)). The point $z_3 = 0$ refers to the point where the cantilever is in contact with the sample but is not bent. A similar curve was also reported in [7], but was not interpreted or analysed with regard to the water film thickness.

A detailed analysis of the interaction between tip and sample is not important for the water film thickness measurement. Therefore, it is only necessary to consider the repulsive interaction, the capillary force and the van-der-Waals-interaction [13, 14].

The distance between the point where the amplitude starts to reduce and the no-bending point of the cantilever in contact gives the distance $z_1$ (see Fig. 2). An analysis of this distance shows for a cantilever amplitude $a_0$=30 nm an increasing value of $z_1$ during reduction of the water vapour pressure (see Fig. 3). A further analysis gives a changing of slope for the bending of cantilever in contact to the sample for different residual gas pressures. Both effects are related to the water film thickness and the adhesion between cantilever tip and sample. An unhindered movement is observed without the presence of water. The cantilever sticks on the surface for thicker water films. For subtracting such effects, the distance $z_1$ is measured for different amplitudes $a_0$. An extrapolation to zero amplitude gives the value $z_1(0)$. Even here, a water film on sample and on tip is observed. An offset for this $z_1(0)$ is measurable without water in ultra high vacuum. This averaged offset was determined to be $z_0$=0.39 nm for silicon oxide and $z_0$=3.97 nm for hydrogen-terminated silicon. The reason for this offset is the direct interaction occurring between the sample and tip. It is not known at this time why the offset values are so different for the two materials. The varying slopes in Fig. 3 are most likely due to the different interactions present between bulk water, double layers and SiO$_2$. Apparently, the softer the material pair, the flatter is the slope of the curve in Figure 3. For purely water-water interactions, the amplitude is not affected because of the inherent flexibility of water molecules. The water film thickness was calculated using the relation,

$$d_{H2O,DDFS} = \frac{z_1(0) - z_0}{2} \tag{2}$$

The interaction between sample and tip starts in air pressure at a distance determined with double of the water film thickness and the half of the vibration amplitude of the cantilever. The distance where the interacting starts is increased at low pressures. The fact that the tip interacts with the sample at separation distances as large as 40 nm (Fig. 3) suggests a long range interaction occurring between the sample and the tip. The nature of this interaction is not known at this point. One possibility is that it could be an electrostatic interaction due to the insulating nature of the native oxide surfaces present on both surfaces. Whatever the exact interaction might be, the analysis with the different amplitudes factors this effect out. A detailed analysis of the interaction occurring between tip and sample is not important for the water film thickness measurement. In this case, it is only necessary to consider the repulsive interaction for the cantilever bending in contact and attractive forces for the reduction of amplitude during the tip-sample approach [13, 14].

## V. RESULTS AND DISCUSSION

The water film thickness was measured as a function of the residual gas pressure by DTS and by DDFS (Fig. 4). Two steps are visible in both curves (Fig 4). The transition from range (a) to (b) is shifted between both measurements, while the transition from range (b) to (c) is in the same pressure range, although also slightly shifted. These shifts occur due to different contact times between the probe tip and sample in DTS and DDFS. In case of DDFS, at first the vibrating cantilever is situated far from the sample (at a distance of approximately 1500 nm). Tip and sample are in contact during measurement of distance by DDFS only for a very short time period. In case of DTS, the probe tip is in constant tunnelling contact also between each of the distance tunnelling spectroscopy measurements. Thus, in the former case, capillary effects exist only during each measurement, while in the latter the capillary neck is also present during the time between the measurements. Additionally electrostatic forces present due to the applied voltage inside the capillary neck [15] tend to stabilise the water in the capillary neck. Since both effects maintain water on the tip, the steps measured by DTS are shifted to lower pressures due to the additionally present electrostatic pressure.

The measured film thicknesses in Table I for DTS measurements are 25 times higher than for DDFS measurement. This phenomenon results from the applied voltage in case of DTS measurements. A pure mechanical contact is used during DDFS, whereas a voltage of 10 V is applied for DTS measurement. This voltage along with the short distance between sample and tip (in the subnanometer range) as well as the field effect for sharp tips leads to a high electric field. The water dipoles align themselves in this field producing an electrical contact long before a mechanical contact occurs [15]. A comparison of both methods shows that the measured values follow the same curve form during reduction of the water partial pressure. Due to this characteristic, it is possible to calibrate DTS values by DDFS measurements. The advantage of doing this lies in the fact that the measurement time for DDFS is much higher because of the need to perform multiple amplitude measurements. Thus, in cases where rapid measurements are needed, like for example during reduction of water partial pressure without thermodynamic equilibrium (see Fig. 4), the higher measuring speed capability of DTS can be implemented to provide more data points without compromising on data accuracy.

Table II presents an overview of water thickness measurements on different materials and under different environmental conditions obtained from literature and this work. The distance of 2 oxygen atoms of different water molecules in an ice crystal is 0.276 nm [19]. An inspection of the values in Table II shows that a monolayer of water molecules has a thickness of (0.25±0.05) nm. Therefore, a film consisting of 2 double layers water molecules is assumed in range (b) of Figure 4. It is important to note that a water double layer is thinner than two monolayers of free water molecules. The measurements of shear force with dynamic shear force spectroscopy as a function of distance [18] (glass-mica-contact) and of the repulsive force during approach in the surface forces apparatus (SFA) [17] (mica-mica-contact), shows that 4 molecular layers are strongly bonded (high shear force, high repulsive force), while a further 4 layers of molecules are weakly bonded (lower shear force, lower repulsive force) in the contact between the two probes. Thereby, gradations of the force are visible in that a double layer of ordered (electrical double layer) and a double layer of semiordered water are present on each of the probes in contact. From these deliberations, models for the water films of about 2.5 nm on $SiO_2$ (tip) in contact against Si(100) (sample) can be established with the following three parts: the water film in range (a) in Fig. 4 consists of a double layer of ordered water molecules (electro-chemical double layer) on the surface of the solid [17, 18, 20]. A transition layer follows this layer with a thickness that is equal to the double layer. The water molecules cannot move freely [17, 18]. Free bulk water forms after this layer [8].

The water film of 0.7 nm thickness (range (b)) consists of an ordered double water layer (electrical double layer) and a transition water double layer. Bulk water desorbs when the water partial pressure of the environment is smaller than the vapour pressure of bulk water (32 mbar at room temperature). The vapour pressure for the ordered and the transition layer was determined from the desorption of this layer to about $10^{-7}$ mbar (at room temperature, see Fig. 4).

For confirmation of this method the water film thickness was measured by DDFS on hydrophobic H:Si(100) (see Fig. 5). Here, no changes of water film thickness were evident. The uncertainty of measurement was determined to smaller than 0.2 nm. This shows a hydrophobic surface without water.

Coming back to the micro devices: the friction force measured by contact scanning force microscopy (SFM) is shown in Fig. 6 during increasing water partial pressure by evacuating a vacuum chamber [10]. Here also three ranges are visible. By comparing the water film thickness measurements and friction force, a correlation becomes possible. The shift of transition from range (a) to (b) is caused by different contact times, as explained above. The tip of the SFM is always in contact with the sample during measurement and water is held in the capillary neck between tip and surface due to the Laplace pressure [21].

## VI. CONCLUSION

Distance dynamic force spectroscopy (DDFS) is a method for measuring the water film thickness from ambient pressure to ultra high vacuum ($10^{-9}$ mbar). Due to the distance spectroscopy the thickness measurement is suitable for continuous water films in contrast to water island microscopy. The method was used for measuring the water film thickness on Si(100) during reduction of the water partial pressure by pump down in a vacuum chamber.

A 2.6±0.2 nm thick water film exists under ambient conditions. The water film is reduced to a thickness of 0.7±0.1 nm between $10^{1}$ and $10^{-6}$ mbar. Two double layers of ordered water exist on the surface, the first nearest to the surface is strongly bonded while the other is only weakly bonded. No detectable water is present on the surface at pressures lower then $10^{-7}$ mbar.

Different advantages exist for distance tunnelling spectroscopy (DTS) and for distance dynamic force spectroscopy (DDFS). A simple current measurement in spectroscopy mode of scanning tunnelling microscopy is used. A comparison between DTS and DDFS shows that DTS results in thickness measurements are 25 times higher. Clearly, DDFS is the more accurate method, although it is time consuming. Therefore, DTS can still be used if calibrated by the more accurate DDFS method.

A correlation of friction force to water film thickness is shown. The described measurement of water film thickness provides a way of characterizing thin water films present on micro devices where surface wettability is an important issue.

## ACKNOWLEDGMENTS

This work was supported by a grant from the Deutsche Forschungsgemeinschaft (Projekt Sche 425/2-4).

TABLES

TABLE I: Measured water film thicknesses by distance tunnelling spectroscopy and distance dynamic force spectroscopy (Fig. 4).

| Water film thickness [nm] | (a) | (b) | (c) |
|---|---|---|---|
| measured by DTS | 66 ± 6 | 17 ± 5 | 0 ± 4 |
| measured by DDFS | 2.6 ± 0.2 | 0.7 ± 0.1 | 0.0 ± 0.1 |

TABLE II: Measured water film thicknesses in comparison to other methods from literature.

|  | Environmental conditions | Materials | Water film thickness [nm] | Structural models for the water films | |
|---|---|---|---|---|---|
| Mono layer range | $10^1 - 10^{-6}$ mbar | $SiO_2$ | 0.7±0.1 | 2 double layer | this work |
|  | electrolyte | silver | >0.3 | 1 double layer | [16] |
|  | air | Gold | 0.2 | 1 mono layer | [7] |
|  | air | Mica | 0.25±0.05 | 1 mono layer | [8] |
|  | air | Mica | 0.25±0.03 | 1 mono layer | [17] |
|  | air | Mica | 0.254 | 1 mono layer | [18] |
|  | air | Mica | 0.85 | 4 mono layers | [18] |
| Multi layer range | air | $SiO_2$ | 2.6±0.2 |  | this work |
|  | air | graphite | 5 |  | [7] |
| Methods: | this work | Distance dynamic force spectroscopy | | | |
|  | [16] | Distance tunnelling spectroscopy | | | |
|  | [7] | Dynamic force microscopy | | | |
|  | [8] | Polarization force microscopy | | | |
|  | [17] | Distance repulsive force spectroscopy | | | |
|  | [18] | Distance dynamic shear force spectroscopy | | | |

FIGURE CAPTIONS

FIG. 1: The tunnelling current $I$ versus the distance $z$ for measuring the water film thickness during approach of tip towards the sample. The schematics above show the interactions occurring between tip, water film and sample.

FIG. 2: The cantilever deflection of a vibrating cantilever as a function of the tip-sample distance $z$ for measuring the water film thickness during approach of the tip towards the sample. The value, $a_0$, is the stimulated amplitude. As in Fig. 1, the upper schematic shows the interactions occurring between tip, water film and sample.

FIG. 3: An example from a typical set of measurements. The diagram shows the distance of the beginning amplitude reducing distance $z_1$ of the vibrating cantilever for different stimulated amplitudes $a_0$ during approach of tip and sample for various residual gas pressures. Additionally, the extrapolated distance of the beginning amplitude reduction for zero amplitude $z_1(a_0=0)$ and the water film thickness for these measurements are given.

FIG. 4: Water film thickness $d$ on hydrophilic $SiO_2$:Si(100) as function of the residual gas pressure, $p$, measured by distance tunnelling spectroscopy (DTS) (top) and by distance dynamic force spectroscopy (DDFS) (bottom) during reduction of the water partial pressure, $p_{H2O}$. The grey bars indicate the uncertainty present in the measurement.

FIG. 5: Water film thickness $d$ on hydrophobic H:Si(100) as function of residual gas pressure $p$ by distance dynamic force spectroscopy (DDFS) during decreasing of water partial pressure $p_{H2O}$. The grey stripe shows the uncertainty of measurement.

FIG. 6: Friction force $F_R$ on hydrophilic $SiO_2$:Si(100) as function of residual gas pressure, $p$, measured by friction force microscopy (FFM) during reduction of the water partial pressure, $p_{H2O}$. The accompanying grey bar shows the uncertainty of the measurement, which is smaller then the symbols in range (c). The used normal load is $F_L$=60 nN by a scanning velocity $v$=300 nm/s and moving distance of $s$=300 nm [10].

Figure 1

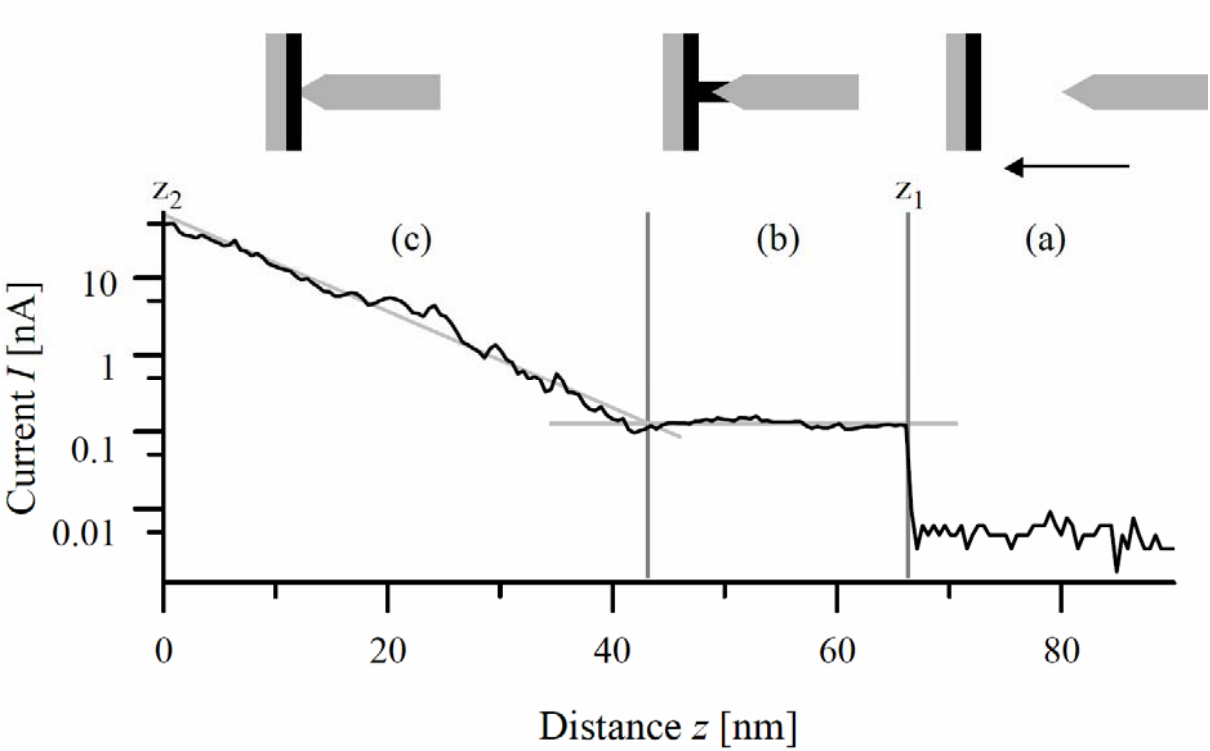

Figure 2

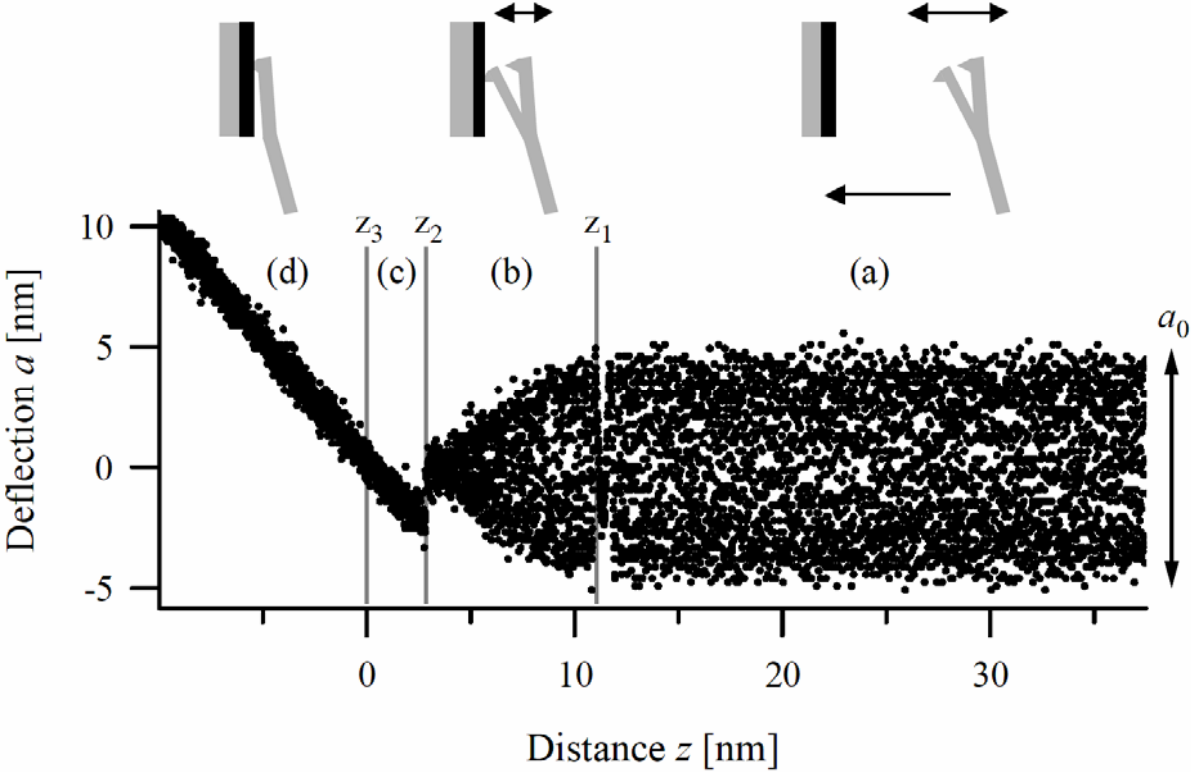

Figure 3

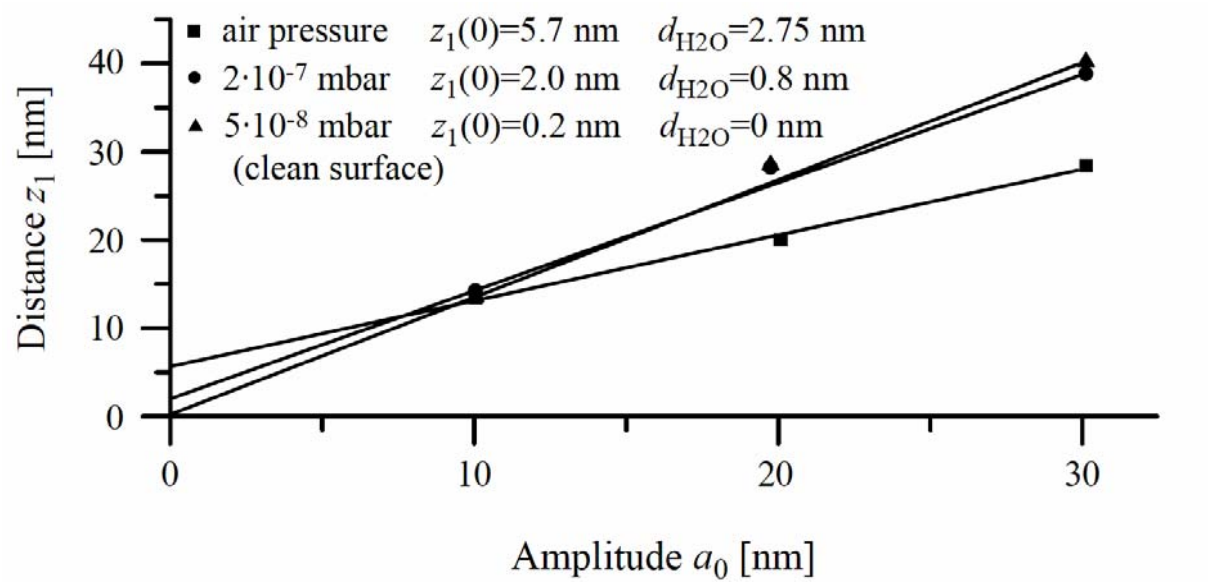

Figure 4

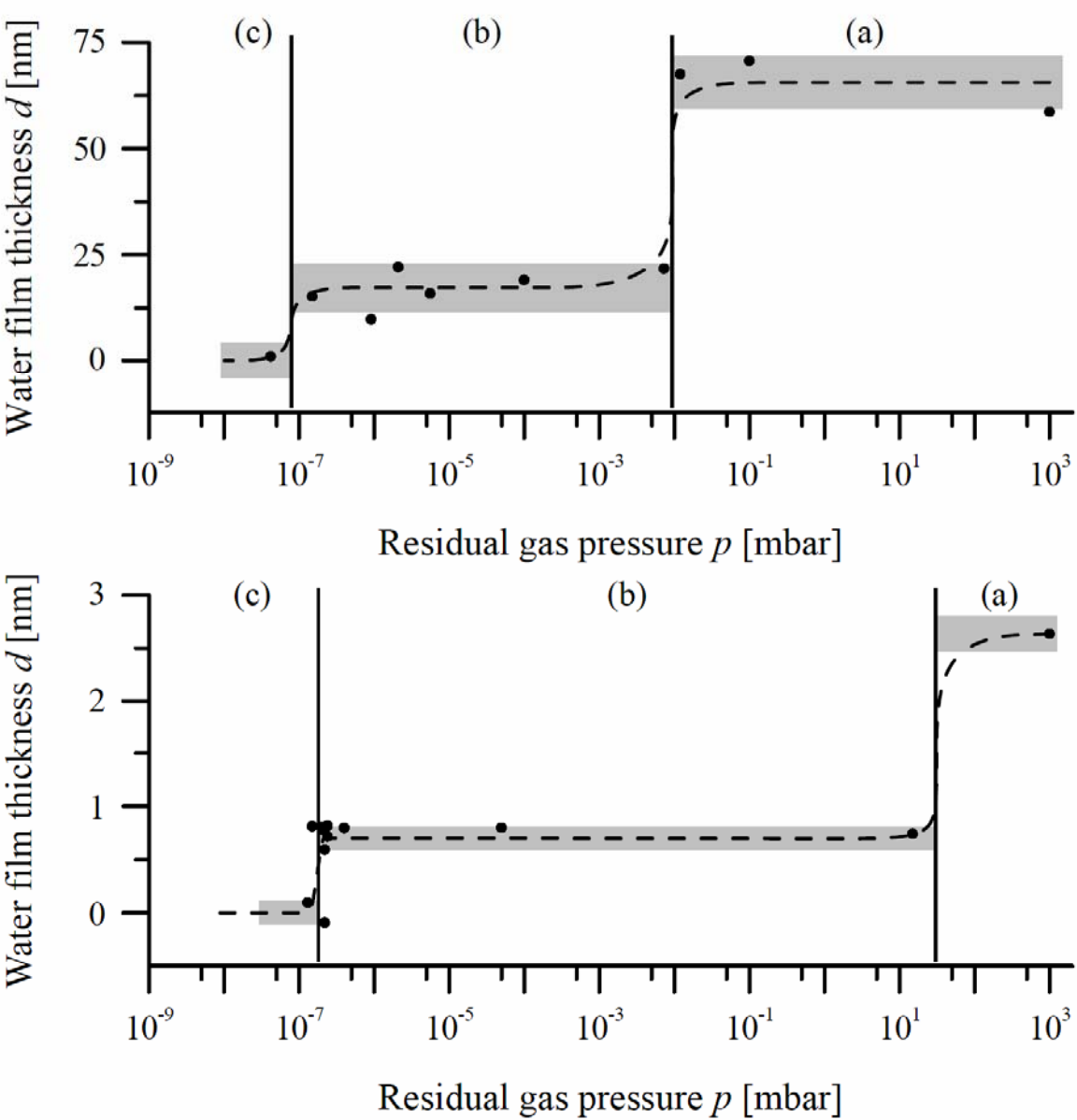

Figure 5

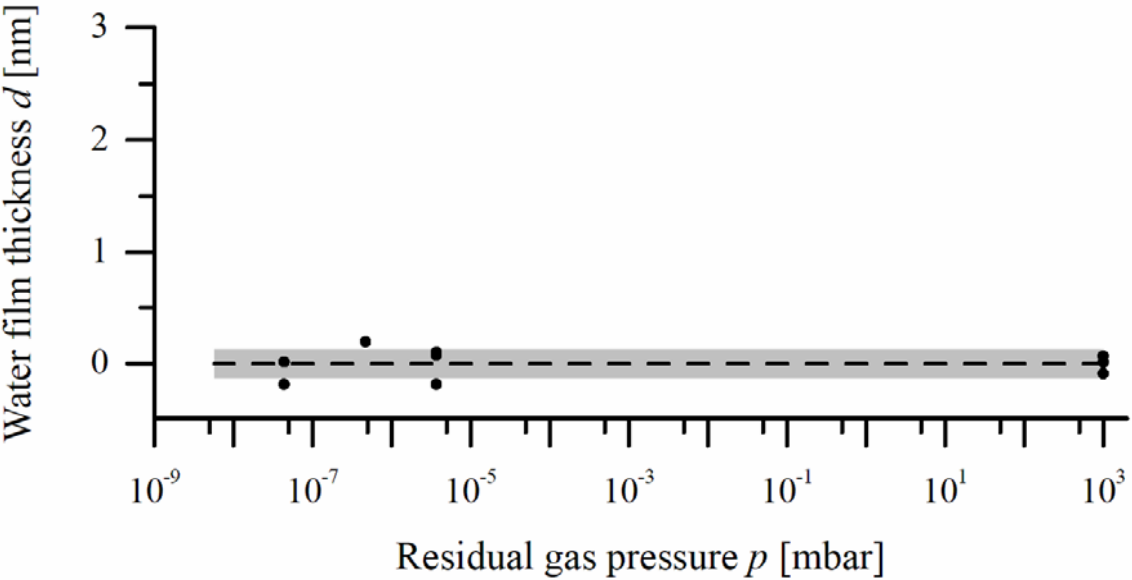

Figure 6

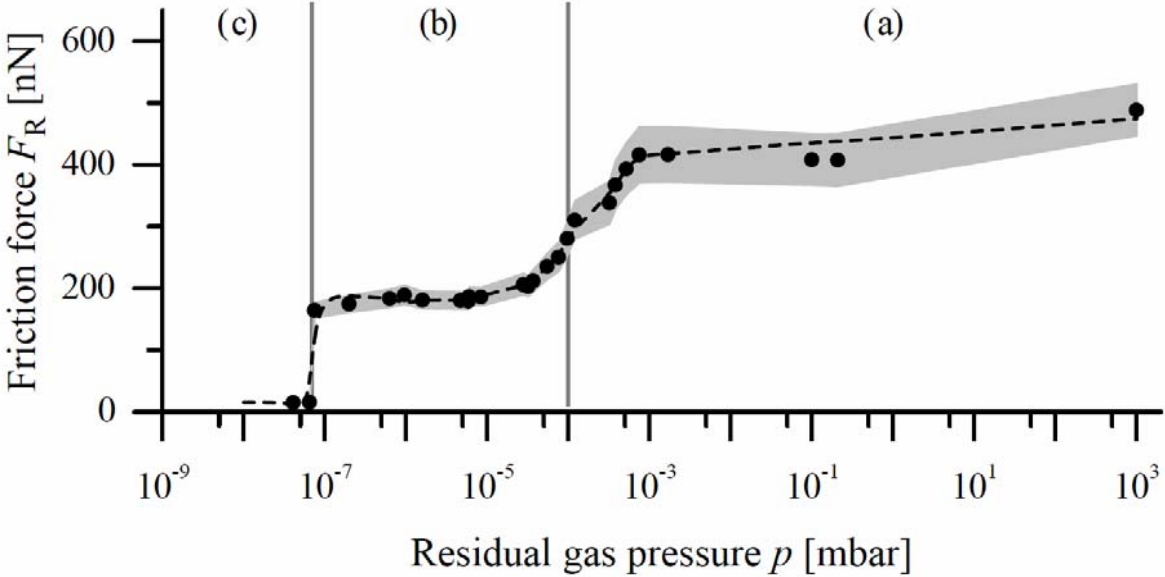